\documentclass[aps,prl,twocolumn,showpacs,superscriptaddress,groupedaddress]{revtex4}  

\usepackage{graphicx}  
\usepackage{dcolumn}   
\usepackage{pbox}
\usepackage{bm}        
\usepackage{amssymb}   
\usepackage{amsmath}
\usepackage{color}
\usepackage{braket}
\newcommand{\comment}[1]{}

\begin{document}
\renewcommand{\arraystretch}{1.5}
\mathchardef\mhyphen="2D

\title{Efficient sideband cooling protocol for long trapped-ion chains}

\author{J.-S. Chen}
\email[Electronic address: ]{chen@ionq.co}
\author{K. Wright}
\author{N.C. Pisenti}
\author{D. Murphy}
\author{K.M. Beck}
\author{K. Landsman}
\author{J.M. Amini}
\author{Y. Nam}

\affiliation{IonQ, Inc., College Park, Maryland, 20740, USA}

\date{\today}

\begin{abstract}
Trapped ions are a promising candidate  for large scale quantum computation. Several systems have been built in both academic and industrial settings to implement modestly-sized quantum algorithms.
Efficient cooling of the motional degrees of freedom is a key requirement for high-fidelity quantum operations using trapped ions.
Here, we present a technique whereby individual ions are used to cool individual motional modes in parallel, reducing the time required to bring an ion chain to its motional ground state.
We demonstrate this technique experimentally and develop a model to understand the efficiency of our parallel sideband cooling technique compared to more traditional methods.
This technique is applicable to any system using resolved sideband cooling of co-trapped atomic species and only requires individual addressing of the trapped particles.
\end{abstract}


\maketitle

Trapped ion quantum information processors are a promising candidate for general purpose quantum computation and small programmable ion-trap quantum computers have been demonstrated in both academia and industry~\cite{monz2016realization,linke2017experimental,wright2019benchmarking}. 
In these processors, quantum information is encoded in the atomic states of a chain of trapped ions, where the collective (secular) motion of the chain acts as a quantum information bus for entangling operations.
To achieve high fidelity quantum operations, it is necessary to prepare the motion in a desired quantum state.
This process generally begins with Doppler cooling using a dipole-allowed transition, followed by various sub-Doppler cooling techniques to drive the system to the motional ground state~\cite{wineland1998}.
Many sub-Doppler cooling methods have been developed to achieve this goal~\cite{diedrich1989laser,roos2000experimental,manfredi2012adiabatic,poulsen2012adiabatic,lin2013sympathetic,ejtemaee20173d}, among which resolved sideband cooling (SBC) is the most general because it does not depend on the atomic structure and can be used to cool all directions of the ions' motion~\cite{monroe1995resolved,chen2017sympathetic}.
While this technique has found powerful demonstrations ranging from atomic clocks to small-scale quantum computers~\cite{linke2017experimental,chen2017sympathetic}, it remains a practical challenge to cool a long chain of ions to the motional ground state. 
The primary difficulty arises from the fact that the required number of cooling pulses, or equivalently the cooling time, scales linearly with the number of secular modes of interest, and many heating mechanisms~\cite{chen2017sympathetic} may become more significant in a longer ion chain. This eventually limits the achievable temperature after sideband cooling.

Individual addressability of qubits~\cite{naegerl1999addressing} has become one of the essential features in recent trapped-ion quantum information processors, which generally utilize an array of tightly focused laser beams separated by a few micrometers to manipulate the qubit state of single ions.
These beams can be steered by acousto(electro)-optic deflectors or micro-electromechanical systems~\cite{schindler2013quantum,crain2014individual,Nigg2016thesis,debnath2016demonstration}, which enable the individual control of a single ion's atomic state by either driving the qubit transition directly or modifying the qubit energy levels via the light shift, without touching the neighboring ions. 
In this letter, we experimentally demonstrate a protocol which uses individual qubit addressability to sideband cool the secular motion of an ion chain in parallel.
This cooling method theoretically allows for ground state cooling with an approximately constant number of pulses as the length of ion chain increases, which is consistent with the experimental demonstrations up to 25 ions.
We develop a simple model of this process and show that it provides a quadratic speedup in the time required for sideband cooling an $N$-ion chain. 
Our protocol is immediately applicable to any experiments using resolved sideband cooling with the individual addressability.

The experimental apparatus has been discussed in detail elsewhere~\cite{wright2019benchmarking}, but we highlight the important aspects here for clarity.
A surface-electrode linear Paul trap is used to trap a chain of  $^{171}\text{Yb}^+$ ions, where the qubit is encoded in two electronic ground state hyperfine levels, ${| ^2S_{1/2},\,F=0,\,m_{F}=0 \rangle} \equiv |\downarrow\rangle$ and ${| ^2S_{1/2},\,F=1,\,m_{F}=0 \rangle} \equiv |\uparrow\rangle$, of each ion.
An anharmonic potential along the trap axis is engineered to provide  equal spacing between ions~\cite{home2011normal}.
The individual manipulation of qubits is achieved by driving  two-photon Raman transitions between qubit states using a pulsed laser at $355$ nm. 
The two Raman beams are counter-propagating and perpendicular to the trap axis, which enables manipulation of secular motion with non-zero projection along the propagation direction of the laser beams.
One of the Raman beams has a broad spatial profile to illuminate all the ions, while the other is split into several beams, each of which is tightly focused and interacts with a single ion.
Using the trap electrodes, we rotate the normal mode coordinate of the ion chain about the trap axis such that only one set of $N$ transverse modes are addressable efficiently by our Raman beam configuration for a chain of $N$ ions.

Resolved sideband cooling of a given mode $m$ involves tuning the Raman lasers near resonance with a red sideband (RSB) transition.
In this regime, we have an effective Hamiltonian of the form
\begin{align}
    H^m_\text{RSB} &= \Omega\eta^m_j\left(\sigma^+_j a_m + \sigma^-_j a^\dag_m \right)\,,
\end{align}
where $\eta^m_j \equiv \xi^m_j \Delta k \sqrt{\hbar/2 M \omega_m}$ is the Lamb-Dicke parameter for ion $j$ and motional mode $m$ with mode vector $\xi^m_j$, $\Omega$ is the two-photon Rabi frequency driving the carrier transition $|\uparrow\rangle \leftrightarrow |\downarrow\rangle$, $\sigma^{+(-)}_j$ is the raising (lowering) operator for qubit $j$, $\Delta k$ is the differential wavevector of the two Raman laser beams, and $a^\dag_m$ ($a_m$) is the creation (annihilation) operator for phonons in motional mode $m$.
The Hamiltonian is turned on for a duration $\tau$ such that the state evolution under $ H^m_\text{RSB}$ results in a transition from $ |\downarrow, n_m\rangle$ to $|\uparrow, (n_m-1)\rangle$, where $|s,n\rangle$ denotes the composite state of the qubit state $|s\rangle$ and the motional mode state $|n\rangle$ with the phonon number $n$.
A short repumping pulse resets the qubit to $|\downarrow\rangle$ while leaving the motion unchanged.
This process is repeated until the mode $m$ has been brought to its motional ground state.

Traditionally, this sideband cooling process is carried out sequentially for each mode $m$, because each mode requires a different resonance frequency. 
In systems where individual ions can be driven with individually controlled Raman beam frequencies, this process can be parallelized by driving each ion with a separate frequency.
What remains is to choose an efficient ion $\Longleftrightarrow$ mode mapping to maximize the set of $\{\eta^m_j\}$, which determines the timescale $\tau$ to complete a RSB transition.
Our scheme, in effect, removes the inefficiency incurred in the traditional sideband cooling scheme, where inevitably some $\eta_j^m$'s are small and thus the cooling rate of mode $m$ by ion $j$ is negligible.
We find experimentally that a suitable mapping always seems possible, rendering a parallel sideband cooling scheme with better efficiency than the traditional non-parallel scheme.

\begin{figure}[t]
\includegraphics[width=\columnwidth]{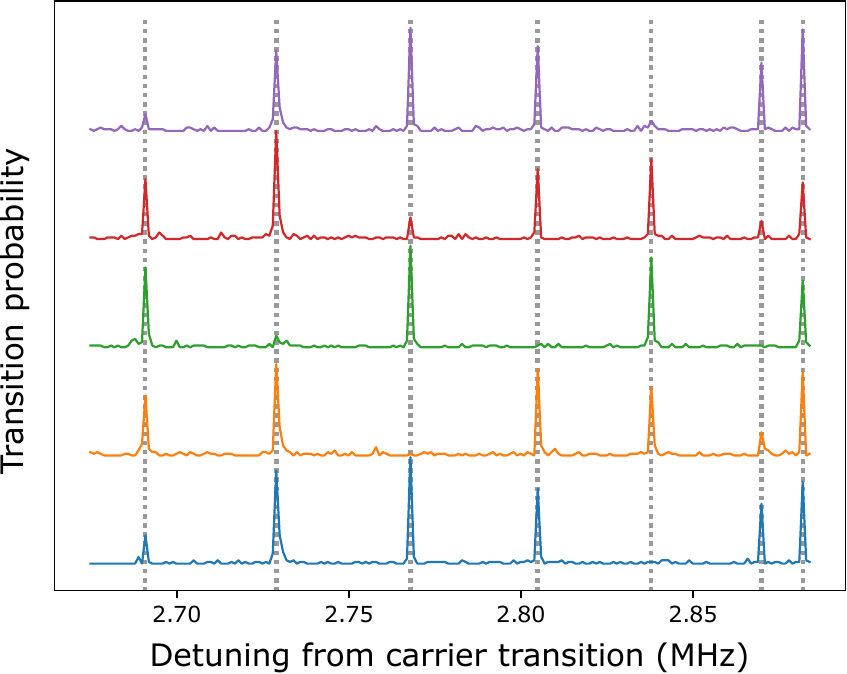}
\caption{Measurement of the different participation of individual ions in a chain to a given mode. Different traces represent the motional spectra measured by different ions, while the vertical dotted lines indicate the frequencies of motional modes detuned from the resonant carrier transition. The traces are offset vertically for visibility.}
\label{fig:participation}
\end{figure}

To illustrate our scheme, we present a realistic example using a seven-ion chain where the middle five ions are evenly spaced in an engineered anharmonic potential well, and the two end ions serve to maintain equal spacing of the middle ions.
The middle set of five ions serve as qubits in a computation, and are individually addressed by the Raman laser system.
The cooling sequence, similar to other trapped-ion systems, starts with a few milliseconds of Doppler cooling, which initializes the ions' collective modes of motion of interest close to the Doppler limit, corresponding to about five motional quanta in each transverse motional mode for $^{171}\text{Yb}^+$ in our experimental condition.
Resolved sideband cooling proceeds using the center five addressable ions to cool all seven modes of interest (namely, one set of transverse modes). Figure~\ref{fig:participation} shows the mode participations for these ions, whose transition strength is proportional  to the magnitude of the secular motion mode vector $\xi^m_j$ belonging to a specific ion.
Each ion couples differently to each motional mode. 
Hence it is essential to choose ions with strong couplings to a particular mode during the sideband cooling process. An example ion-mode mapping is provided in the supplemental information for clarity.

Because there is great flexibility in our mapping between the cooling ion $j$ and secular mode $m$, we can apply more cooling pulses to modes with a higher heating rate without  much increase to the total cooling time~\cite{james1998quantumdynamics}. 
We typically choose to apply two to three sideband cooling sequences, where each sequence cools a predefined set of modes using a subset of the addressable ions. We determine the composition of a given sequence to improve cooling efficiency on higher heating modes, where each cooling sequence has at least one ion cooling the highest heating mode.
For transverse modes, these correspond to the modes at higher frequencies. 
Note again that the mapping between the ions and the modes can be different in different sequences to improve the cooling efficiency. 
In our experiments, we apply pulses with varying duration corresponding different Fock states~\cite{chen2017sympathetic}, although using cooling pulses with the appropriately chosen fixed duration has little impact to the cooling efficiency~\cite{che2017efficientcooling}. 
Generally we need to apply approximately 100 pulses in total to produce motional states cold enough for high-fidelity quantum computation. 

\begin{figure*}
\includegraphics[width=17.5cm]{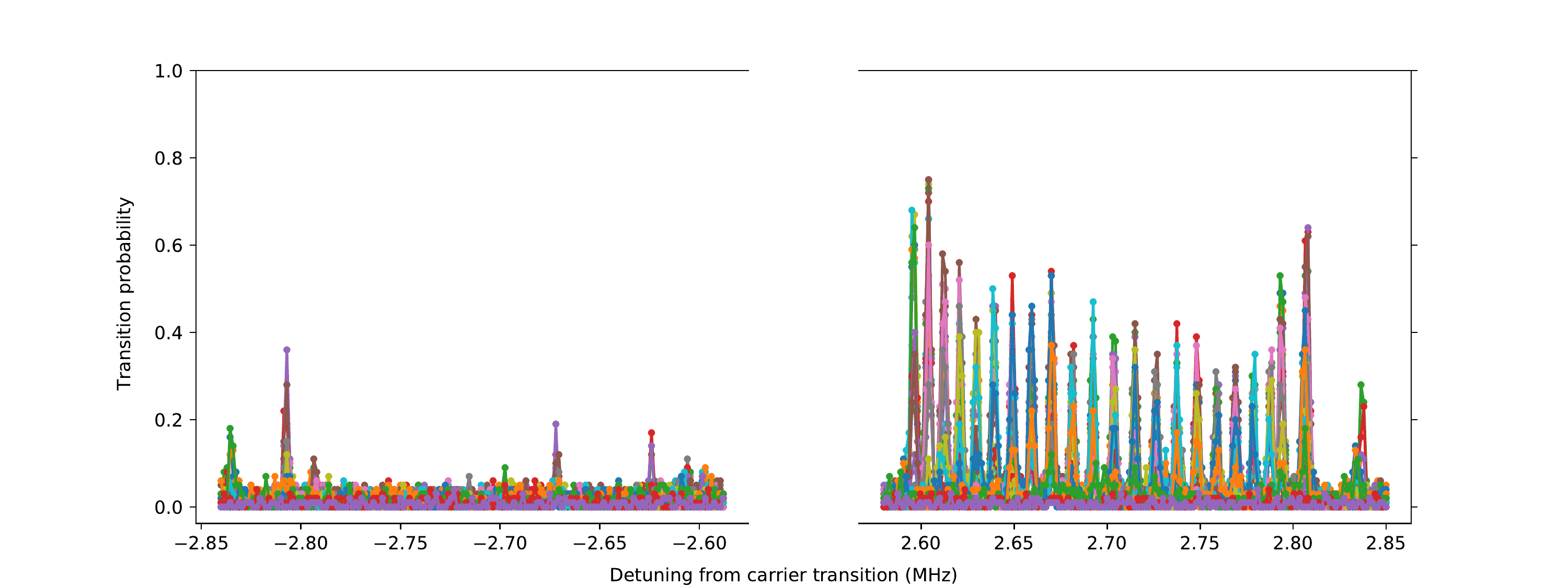}
\caption{Motional sideband spectrum of a $25$-ion chain after  parallel sideband cooling. Note that only 21 ions are addressable while 25 modes are cooled as detailed in the main text. Different traces represent the measurement of different ions. All the modes of motions are cooled to below $0.5$ quanta after sideband cooling.}
\label{fig:asym}
\end{figure*}

We design cooling sequences based on the heuristic detailed above to cool ion chains with a varying number of ions ranging from five up to 25 ions.
While the number of ions grows, the number of cooling pulses needed remains nearly constant using the parallel scheme and thus the time is the same if the pulse duration is the same (see Supplemental Material). 
In Figure \ref{fig:asym}, we show the sideband asymmetry of transverse modes in a 25-ion chain after applying approximately 100 parallel sideband cooling pulses, which takes about 10-20~ms.
While most of the modes have an average phonon number well below $0.2$ quanta, there are four modes that only reach about $0.5$ quanta. We estimate that this is limited by the heating rate, which is about $300\,\text{quanta}/\text{s}$ for a single ion. Practically, the parallel sideband cooling can be optimized for higher two-qubit entangling gate fidelity by applying more cooling pulses on the modes that contribute most to the enclosed geometric phase~\cite{zhu2006trapped,leung2018robust} without additional overhead.

To understand the cooling efficiency compared to traditional, non-parallel SBC methods, we develop a simple model of the cooling process.
Our goal is to find a one-to-one mapping between the set of $N$ ions and the set of $N$ modes, such that for each ion-mode pair $(j,m)$ in this mapping, a non-negligible $\eta^{m}_{j}$ results. 
For an evenly-spaced linear chain of $N$ ions, the change in potential energy of each ion due to a transverse displacement from equilibrium may be treated to second order.
Assuming the interaction strengths between non-neighboring ions are small compared to that of the neighboring ions, the normal mode vectors are then well-characterized by the harmonics of an open pipe. 
Specifically, the structure of $\eta$'s, proportional to the normal mode vectors, may be written as
\begin{equation}
\label{eta_cos}
\eta^m_j = \eta \cos\left(\frac{mj\pi}{N-1}\right),
\end{equation}
where $\eta$ is a constant, $j$ indexes the ion, and $m$ indexes the mode. 
From this structure, we can deduce that for a given mode $m$ there exists at least one antinode $j_m$ such that $j_m$ is not an antinode of any modes other than $m$, with the exception of the center-of-mass mode, where any choice of $j_0$ suffices.

For simplicity, we assume the chain is initialized in a pure Fock state with $M$ phonons per mode.
For the traditional SBC scheme, since all pulses must be applied sequentially, we require $N_{pulse} = N M$ pulses.
In comparison, for our parallel scheme, the number of pulses does not increase with the number of ions but depends only on the number of phonons $M$.
In terms of time, as detailed in the Supplementary Material, our parallel cooling scheme  offers a quadratic advantage over the traditional cooling scheme.
Assuming the approximation made in (\ref{eta_cos}) is valid, it is straightforward to show that the traditional cooling time $\mathcal{T}_\text{cool}^\text{trad}$ of $M$ phonons in each mode is
\begin{align}
         \mathcal{T}_\text{cool}^\text{trad} = \sum_{m = 0}^{N - 1}\left(\sum_{j = 0}^{N - 1}{\cos^2\left(\frac{mj\pi}{N-1}\right)}\right)^{-\frac{1}{2}} \left(\frac{\pi}{2\Omega\eta}\sum_{n = 1}^{M}{\frac{1}{\sqrt{n}}}\right),
    \label{eq:trad_cool_time}
\end{align}
which may be compared to the parallel cooling time $\mathcal{T}_\text{cool}^\text{para}$ of $M$ phonons in each mode
\begin{equation}
\mathcal{T}_\text{cool}^\text{para} = \max_m\left\{\left|\cos\left(\frac{m j_m\pi}{N - 1}\right)\right|^{-1}\right\} \left(\frac{\pi}{2\Omega\eta}\sum_{n = 1}^{M}{\frac{1}{\sqrt{n}}}\right),
\label{eq:para_cool_time}
\end{equation}
where $j_m$ is an appropriate choice of ion $j$ to cool mode $m$ such that $\eta_{j_m}^m$ is large to achieve efficient cooling of that mode.

\begin{figure}[t]
    \includegraphics[width=\columnwidth]{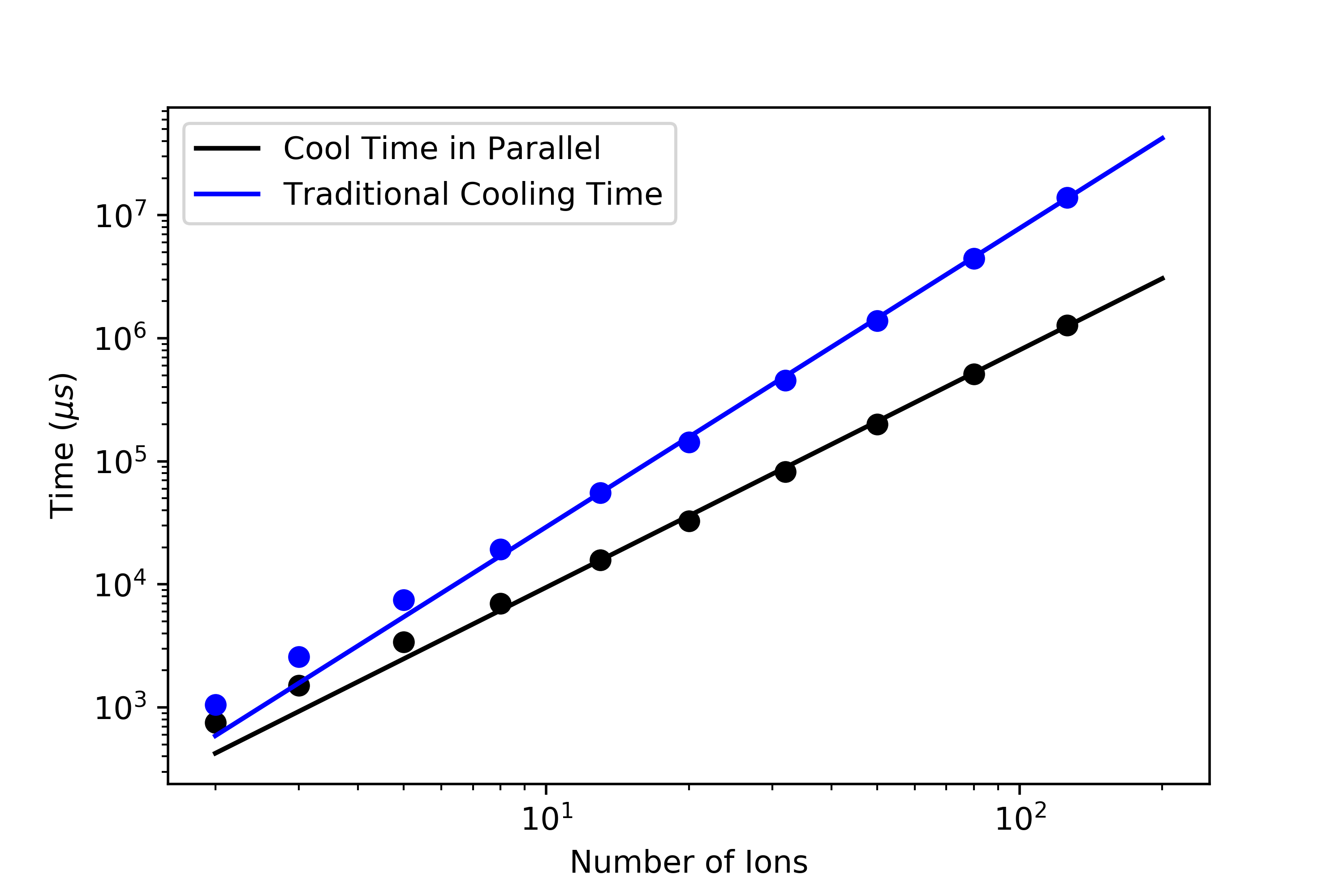}
    \caption{Numerical simulation of time scaling for traditional and parallel cooling techniques in an anharmonic linear Paul trap. The ion spacing is set at $4~\mu $m. Blue: traditional cooling; black: parallel cooling. We chose $\Omega$ such that the rotating wave approximation used to construct the model Hamiltonian that describes the system remains in its validity limit. See Supplemental Material for more detail. The cooling time ratio scales approximately as $N^{1/2}$.
    }
    \label{fig:scaling}
\end{figure}

Figure~\ref{fig:scaling} shows the simulation results of cooling times using (\ref{eq:trad_cool_time}) and (\ref{eq:para_cool_time}) for the number of ions $N = 2, .., 200$.
To verify our experimental results, we chose to fix the ion spacing at $4\,\mu$m in the simulation.
We observe that the parallel cooling time gives a speedup of $N^{1/2}$ over the traditional methods in the asymptotic limit, although deviations are expectedly observed at small $N$.
While the pure Fock state is certainly not the precise state we anticipate after applying the Doppler cooling to a chain of ions, we believe the model we used captures the essence of the efficiency improvement of our parallel approach over the traditional approach.

In summary, we present an efficient  sideband cooling scheme which leverages individual addressability to cool motional modes in parallel.
We demonstrate it on an ion chain consisting of up to $25$ ions, and estimate that our parallel cooling scheme provides a quadratic improvement in cooling time compared to traditional sideband cooling schemes.
The techniques presented here are immediately applicable to any ion-trap quantum information experiment with individual qubit control, without the need for additional lasers or modifications to existing hardware.
Furthermore, it can be used to cool any system with resolved sidebands and individual addressability more efficiently.
The improvement in sideband cooling efficiency is essential for trapped-ion quantum computing architectures which require long chains of qubits.
Extensions of this scheme, for example in combination with continuous sideband cooling techniques~\cite{Jurcevic2017thesis} and higher order SBC~\cite{wan2015efficient,chen2017sympathetic,che2017efficientcooling}, could further reduce the cooling time and may thus be of interest for future work.

%


\renewcommand\theequation{S\arabic{equation}}    
\setcounter{equation}{0}

\section*{Supplemental Material}
\subsection{Example parallel sideband cooling sequence}
As described in the main text, our parallel sideband cooling starts with a few milliseconds of Doppler cooling to prepare the ions' motion close to Doppler limit temperature followed by optical pumping to initialize all the ions to the state $|\downarrow\rangle$.
We can then cool one set of 7 transverse modes ($m_i,~i=0...6$) on a 7-ion chain with five qubits (addressable ions; $q_i,~i=0...4$) using two rounds of parallel sideband cooling. 

In the first round, we apply pulses where the qubits $q_0$ to $q_4$ are driven at the frequencies corresponding, respectively, to resonance for modes $[m_3, m_0, m_2, m_6, m_5]$. The mode participation of each of the qubits is shown in Fig.~1. Mode 6 ($m_6$) has the smallest detuning and the most participation for $q_2$, while mode 0 ($m_0$) has the largest detuning, nearly equal participation from all qubits and, typically, the highest heating rate. We apply 40 of these pulses with increasing duration, timed to the transition time for decreasing phonon Fock states in the specified mode driven by the specified qubit. In between each sideband cooling pulse, we optically pump the ions to state $|\downarrow\rangle$.

The second round is the same as the first round using the qubit-mode mapping $[m_5,m_0,m_4,m_2,m_1]$.
Note that the modes $m_0$, $m_2$, and $m_5$ appear twice in this mapping to achieve the better cooling results.

\subsection{Theoretical model}
Here, we discuss our model for comparing the traditional (sequential) resolved sideband cooling to our parallel scheme.
Specifically, we consider the time each method takes to cool the Fock state $|m=M\rangle^{\otimes N}$ of an $N$-ion chain. 

\begin{figure}
    \includegraphics[width = 8.6cm]{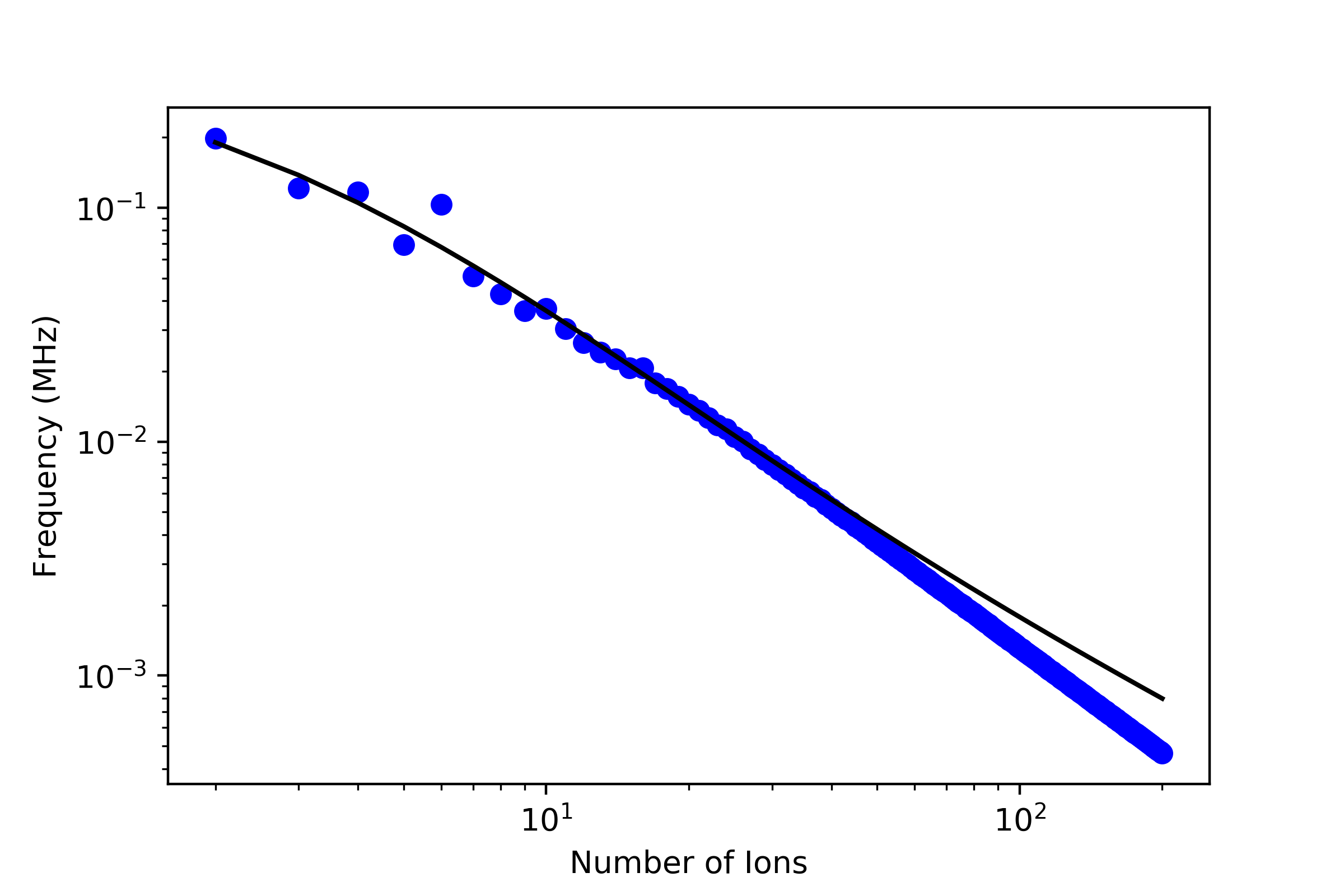}
    \caption{Scaling of the applied carrier Rabi frequency, as a function of the ion number $n$, to remain in the validity limit of the model Hamiltonian in \ref{eq:Trad_Ham}, which sets the motional Rabi frequency to be less than $10\%$ of any two adjacent modes. Considered here is a typical case of a $4~\mu m$-spaced ion chain with $\omega_m \in [2.6,2.9]$ MHz.}
    \label{fig:frequencyscaling}
\end{figure}

Resolved sideband cooling works by transferring individual motional mode excitations to the spin state of an ion, which is reset via optical pumping.
Traditionally, this is done by addressing the whole chain of ions to cool one mode at a time.  In particular, for every motional mode $m$, we drive the spin transition of all of the ions with a detuning $\delta$ such that $\delta - \omega_m = 0$, where $\omega_m$ denotes the motional mode frequency of the mode $m$. More concretely, the traditional, resolved, red-sideband Hamiltonian $\mathcal{H}^m_{\textrm{RSB}}$ for cooling the $m$th mode of an $N$-ion chain is
\begin{equation}
\mathcal{H}^m_{\textrm{RSB,trad}} = \sum_ {j = 0}^{N-1}{\Omega\eta^{m}_{j}\sigma^j_-a^\dag_m + h.c.},
\label{eq:Trad_Ham}
\end{equation}
where $\Omega$ is the Rabi rate of the transition, which we assumed to be uniform across all of the ions for convenience, $\eta_j^m$ is the Lamb-Dicke parameter between ion $j$ and mode $m$, $\sigma^j_{-}$ is the spin lowering operator for $j$th ion, $a^\dag_m$ is the phonon creation operator for mode $m$, and $h.c.$ denotes Hermitian conjugate. The Hamiltonian in (\ref{eq:Trad_Ham}) is valid if, for all ions $j$, the mode-frequency difference between any chosen pair of modes, e.g., $\omega_{m} - \omega_{m'}$, is much greater than $\Omega_j\eta^j_{m'}$, since in this limit (see Fig.~\ref{fig:frequencyscaling} for an appropriate choice of $\Omega$ for a typical case of ion chain) we can rotate away all modes but one and consider the one remaining mode in the Hamiltonian as in (\ref{eq:Trad_Ham}).

Following from the Hamiltonian in (\ref{eq:Trad_Ham}), we can write the time $t^M_{m,\textrm{trad}}$ required to cool $M$ phonons in mode $m$ as
\begin{equation}
\label{eq:tTrad}
t^M_{m,\textrm{trad}} = \frac{\pi}{2\Omega\sqrt{\sum_{j=1}^N {\left(\eta^{m}_j\right)^2}}}\sum_{n = 0}^{M-1}\frac{1}{\sqrt{1+n}},
\end{equation}
where we ignore the relatively short time it takes for optical pumping to reset the spin state of the ions.
Equation (\ref{eq:tTrad}) can be derived by considering the evolution of the state
\begin{equation}
\ket{\psi} = \ket{0}_0\ket{0}_1 ... \ket{0}_N-1 \ket{M'}_m,
\end{equation}
where the first $N$ kets denote the spin state of each ion and the last ket denotes the phonon state of mode $m$, forward in time by $t_{m,\textrm{trad}}^{M'\rightarrow M'-1}$, which results in
\begin{align}
&\exp\left(-i \mathcal{H}^m_{\textrm{RSB,trad}} t_{m,\textrm{trad}}^{M'\rightarrow M'-1} \right) \ket{\psi} \nonumber \\
&= \sum_{j=0}^{N-1}\left[ -\frac{i\eta_j^m}{\sqrt{\sum_{l=0}^{N-1} \left(\eta_l^m\right)^2}} \bigotimes_{k=0}^{N-1}\left(\ket{\delta_{j,k}}_k\right) \right] \ket{M'-1}_m,
\end{align}
where $\bigotimes$ denotes the tensor product, $\delta_{j,k}$ is a Kronecker delta, and
\begin{equation}
t_{m,\textrm{trad}}^{M'\rightarrow M'-1} = \frac{\pi}{2\sqrt{M'}\sqrt{\sum_{j=1}^N \left(\eta_j^m \right)^2 }}.
\end{equation}
In the traditional approach, all of the motional modes are cooled in sequence and the total time $\mathcal{T}_\text{cool}^\text{trad}$ to cool all of them is
\begin{equation}
\mathcal{T}_\text{cool}^\text{trad} = \sum_{m=1}^N t^M_m,
\label{eq:Trad_Time_Total}
\end{equation}
where we assumed the initial phonon number of $M$ for each mode for simplicity.

In our parallel approach, we allocate one ion per each mode. Denoting this one-to-one relation between the ions and the modes with the help of the notation $j_m$, where $j$ is the ion index, $m$ is the mode index, and $j_m$ then denotes ion $j$ used to cool mode $m$, our parallel, resolved, red-sideband Hamiltonian $\mathcal{H_{\textrm{RSB,para}}}$ is
\begin{equation}
\mathcal{H_{\textrm{RSB,para}}} = \sum_{m=0}^{N-1} \mathcal{H^\textrm{m}_{\textrm{RSB,para}}},
\label{eq:Para_Ham}
\end{equation}
where
\begin{equation}
\mathcal{H^\textrm{m}_{\textrm{RSB,para}}}=  \Omega\eta^{m}_{j_m}\sigma_-^{j_m} a^\dag_{m} + h.c.
\label{eq:Para_Ham_mode}
\end{equation}
is the $m$th-mode Hamiltonian for our parallel approach.  

From (\ref{eq:Para_Ham_mode}), following similar steps as the ones used in the traditional cooling discussion above, we may compute the time $t^M_{m,\textrm{para}}$ required to cool $M$ phonons in mode $m$. This results in
\begin{equation}
t^M_{m,\textrm{para}} = \frac{\pi}{2\Omega\eta^{m}_{j_m}}\sum_{n = 0}^{M-1}\frac{1}{\sqrt{1+n}}.
\end{equation}
Assuming once again we have $M$ phonons in each mode, the time $\mathcal{T}_\text{cool}^\text{para}$ required to cool the chain to the ground state of all modes is
\begin{equation}
\mathcal{T}_\text{cool}^\text{para} = \textrm{max}_m\left\{t^M_{m,\textrm{para}}\right\}.
\label{eq:Para_Time_Total}
\end{equation}

\begin{figure}
        \includegraphics[width=\columnwidth]{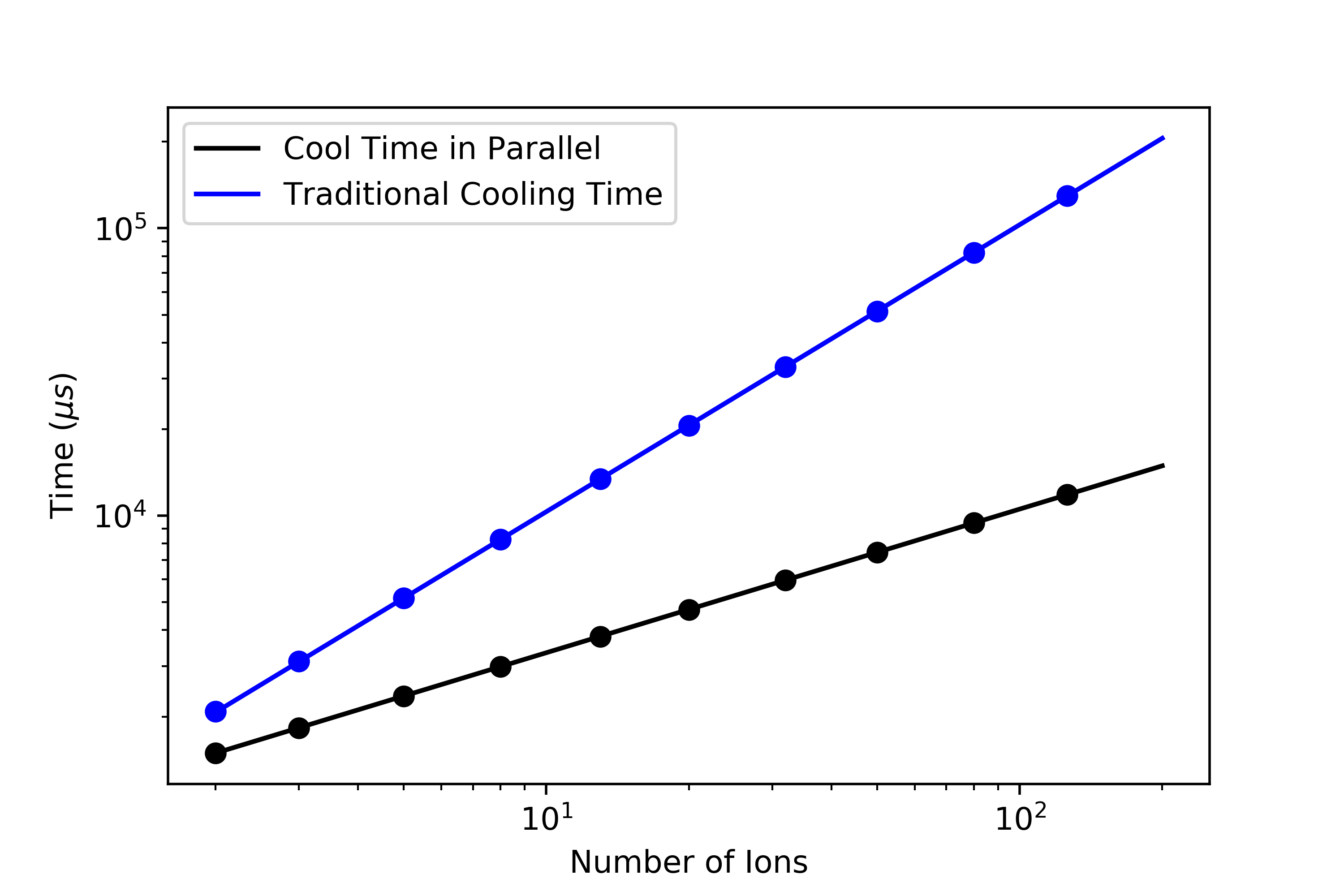}
        \caption{A comparison of cooling time between parallel and traditional methods in a trap with the same motional spectrum at about $2.9$ MHz. The carrier Rabi frequency is independent of the number of ions and set at $0.1$ MHz.}
        \label{fig:scaling2}
\end{figure}

We now compare the two results $\mathcal{T}_\text{cool}^\text{trad}$ in (\ref{eq:Trad_Time_Total}) and $\mathcal{T}_\text{cool}^\text{para}$ in (\ref{eq:Para_Time_Total}). For the analysis to follow, we assume that $\eta^m_j = \eta\cos(m j \pi/(N - 1))$ [see main text Eq.~(1) and discussions around it]. In this case, we have
\begin{align}
         \mathcal{T}_\text{cool}^\text{trad} = \sum_{m = 0}^{N - 1}\left(\sum_{j = 0}^{N - 1}{\cos^2\left(\frac{mj\pi}{N-1}\right)}\right)^{-\frac{1}{2}} \left(\frac{\pi}{2\Omega\eta}\sum_{n = 1}^{M}{\frac{1}{\sqrt{n}}}\right),
\end{align}
which is the main text Eq.~(3), and
\begin{equation}
\mathcal{T}_\text{cool}^\text{para} = \max_m\left\{\left|\cos\left(\frac{m j_m\pi}{N - 1}\right)\right|^{-1}\right\} \left(\frac{\pi}{2\Omega\eta}\sum_{n = 1}^{M}{\frac{1}{\sqrt{n}}}\right),
\end{equation}
which is the main text Eq.~(4).
Figure \ref{fig:scaling2} shows the comparison between $\mathcal{T}_\text{cool}^\text{trad}$ and $\mathcal{T}_\text{cool}^\text{para}$ with $\Omega=0.1$ MHz for various number of ions.
The advantage our parallel approach provides over the traditional approach is clearly shown. The quotient $Q = \mathcal{T}_\text{cool}^\text{trad}/\mathcal{T}_\text{cool}^\text{para}$ is
\begin{equation}
Q = \frac{\sum_{m = 0}^{N - 1}\left(\sum_{j = 0}^{N - 1}{\cos^2\left(\frac{mj\pi}{N-1}\right)}\right)^{-\frac{1}{2}}}
{\max_m\left\{\left|\cos\left(\frac{m j_m\pi}{N - 1}\right)\right|^{-1}\right\}},
\end{equation}
which may be simplified as
\begin{equation}
Q =  \min_m\left|\cos\left(\frac{m j_m\pi}{N - 1}\right)\right| \left( \frac{2}{\sqrt{N}}+\frac{\sqrt{2}(N-2)}{\sqrt{N+1}}\right).
\label{eq:Quotient}
\end{equation}

Equation~(\ref{eq:Quotient}) shows the speed-up offered by our parallel approach scales as $\mathcal{O}(\sqrt{N})$, rendering our parallel approach increasingly better suited for larger systems. This is consistent with our simulation results shown in Fig.~\ref{fig:scaling2}. We also note that the constant in front depends on the mapping between the ions and the modes. A careful choice of the mapping can be made to ensure the constant is large, i.e., the parallel advantage is significant.

\end{document}